\documentclass[runningheads]{llncs}

\usepackage[T1]{fontenc}
\usepackage{graphicx}
\usepackage{booktabs}
\usepackage{amsmath}
\usepackage{CJKutf8}
\usepackage{indentfirst}
\usepackage{cite}
\usepackage{graphicx}
\usepackage{float}
\usepackage{subfigure}
\begin{document}
\begin{CJK}{UTF8}{gbsn}

\title{Text-oriented Modality Reinforcement Network for Multimodal Sentiment Analysis from Unaligned Multimodal Sequences}

\def\CICAISubNumber{273} 
\titlerunning{CICAI 2023} 
\authorrunning{Y. Lei et al.} 
\author{
Yuxuan Lei\inst{1}\and
Dingkang Yang\inst{1} \and
Mingcheng Li\inst{1} \and
Shunli Wang\inst{1}\ \and
Jiawei Chen\inst{1} \and
Lihua Zhang\inst{1,2,3*}}
\institute{
Academy for Engineering and Technology, Fudan University, Shanghai, China\and
Jilin Provincial Key Laboratory of Intelligence Science and Engineering, Changchun, China\and
Engineering Research Center of AI and Robotics, Ministry of Education, Shanghai, China\\
\email{\{yxlei22, mingchengli21, jwchen22\}@m.fudan.edu.cn}\\
\email{\{dkyang20, slwang19, lihuazhang\}@fudan.edu.cn}
}
\maketitle            
\let\thefootnote\relax\footnote{* Corresponding author}
\begin{abstract}
Multimodal Sentiment Analysis (MSA) aims to mine sentiment information from text, visual, and acoustic modalities. Previous works have focused on representation learning and feature fusion strategies. However, most of these efforts ignored the disparity in the semantic richness of different modalities and treated each modality in the same manner. That may lead to strong modalities being neglected and weak modalities being overvalued. Motivated by these observations, we propose a Text-oriented Modality Reinforcement Network (TMRN), which focuses on the dominance of the text modality in MSA. More specifically, we design a Text-Centered Cross-modal Attention (TCCA) module to make full interaction for text/acoustic and text/visual pairs, and a Text-Gated Self-Attention (TGSA) module to guide the self-reinforcement of the other two modalities. Furthermore, we present an adaptive fusion mechanism to decide the proportion of different modalities involved in the fusion process. Finally, we combine the feature matrices into vectors to get the final representation for the downstream tasks. Experimental results show that our TMRN outperforms the state-of-the-art methods on two MSA benchmarks.

\end{abstract}

\section{Introduction}
\noindent Recognizing the research value of sentiments, numerous studies \cite{yang2023context, yang2022emotion, yang2022contextual, yang2023target,du2021learning} in recent years have focused on identifying and analyzing human sentiments. Compared with traditional unimodal sentiment analysis, Multimodal Sentiment Analysis (MSA) attempts to mine sentiment information from multiple data sources to more comprehensively and accurately understand and predict a wide range of complex human emotions. 

While data from multiple modalities can be complementary, the asynchrony between different modality sequences caused the distress of fusion. To address this problem, most prior works have manually aligned visual and acoustic sequences at the resolution of text words~\cite{tsai2018learning, wang2019words}, but this has also resulted in high labor costs and ignored long-term dependencies between different modal elements. Recent efforts like~\cite{tsai2019multimodal, lv2021progressive} have tended to deal with unaligned multimodal sequences by cross-modal attention. They often digest inter-modality correlations through sufficient interactions between each pair of modalities. However, this results in a surge in the number of parameters and redundant information in the modalities.  They treat all modalities with the same weight without regard to the fact that the semantic richness of distinct modalities is different, which may lead to strong modalities being neglected and weak modalities being overvalued. 
Observing previous works~\cite{wu2021text, chen2017multimodal}, we found that text modality dominates the MSA task. On the one hand, the text modality is naturally highly structured and semantically condensed; on the other hand, due to the maturity of natural language processing techniques, modeling techniques for text data are relatively mature. In this situation, it is crucial to balance the contributions of different modalities. Moreover, the vanilla Transformer\cite{vaswani2017attention} also has some drawbacks. The self-attention mechanism incorporates redundancy and noise while focusing on the information within the modality, especially for the visual and acoustic modalities. Unlike spoken words that can be encoded directly, acoustic and visual modalities are pre-processed before being fed into the network, and noise is inevitably introduced during the pre-processing process~\cite{chen2017multimodal}. Secondly, the redundancy in time series between visual and acoustic sequences is very high.  

Inspired by the above observations, we propose a Text-oriented Modality Reinforcement Network (TMRN) to refine multimodal representations effectively. The core strategy of the TMRN is to interact between modalities with the text modality at the center and to guide the reinforcement process of the other two modalities by text modality. For the inter-modal intersection, we propose a text-centered cross-modal attention module to make full interaction for text/acoustic and text/visual pairs.  We also present an adaptive fusion mechanism to measure the weights of the different modalities during fusion. For the intra-modal reinforcement, we design a text-gated self-attention module to introduce prior knowledge of textual semantics in the process of feature reinforcement of visual/acoustic modalities. This aims to mine the semantic information on time series better and to ignore the noise of visual/acoustic modalities. Overall, we make the following three contributions:
\begin{itemize}
    \item We propose TMRN, a method that focuses on the dominance of the text modality in MSA tasks. The TMRN interacts and reinforces the other two modalities with the text modality as the main thread to obtain a low redundancy and denoised feature representation.
    \item We present a Text-Centred Cross-modal Attention (TCCA) module and a Text-Gated Self-Attention (TGSA) module to mine inter-modal and intra-modal contextual relationships.
    \item We perform a comprehensive set of experiments on two human multimodal language benchmarks MOSI~\cite{zadeh2016mosi} and MOSEI~\cite{zadeh2018multimodal}. Our experiments show that our method achieves state-of-the-art methods on these two datasets.
\end{itemize}

\section{Related Work}
\noindent Human multimodal sentiment analysis is to infer human emotional attitudes from the various modality information in video clips. Compared to multimodal fusion from static modalities like images~\cite{mittal2020emoticon}, the key technique for this task is how to fuse time-series sequences from different modalities such as natural language, video frames, and acoustic signals~\cite{tsai2018learning}, especially when these sequences are temporally unaligned. Some recent works~\cite{tsai2018learning,wang2019words} have focused on manually aligning the visual and acoustic sequences in the resolution of textual words before training. However, manual word alignment is costly, and there is inevitably some loss of information in the multimodal fusion after alignment.               

Furthermore, some researchers have worked on unaligned multimodal data. These works can be classified into two categories: discarding the time series dimension and retaining the time series dimension in the subsequent modal interactions. For the former, they usually take one row of the two-dimensional features as a feature vector for subsequent interaction and fusion \cite{hazarika2020misa,yu2021learning,yang2022learning}. \cite{hazarika2020misa,yang2022learning} learned modality-invariant and modality-specific representations to give a comprehensive and disentangled view of the multimodal data. \cite{yu2021learning} jointed training the multimodal and unimodal tasks to learn the consistency and difference, respectively. For the latter, they tend to use the attention mechanism to implement interactions between non-aligned sequences \cite{tsai2019multimodal,lv2021progressive}.

A great deal of attention to attention mechanism has been triggered by the Transformer~\cite{vaswani2017attention}.  Transformer networks have been successfully applied to many tasks like semantic role labeling \cite{strubell2018linguistically} and word sense disambiguation\cite{tang2018self}. And now, Transformer is also widely used in the multimodal field. \cite{tsai2019multimodal} presented Multimodal Transformer (MulT), which uses cross-modal attention to capture the bimodal interactions without manually aligning the three modalities. \cite{lv2021progressive} proposed PMR, which is a further improvement of the interaction between the three modalities based on MulT. Following the latter approaches, our work is also based on the attention mechanism. 
\section{Problem Statement and Model Structure}
\subsection{Problem Statement}
\begin{figure}

\centering 
\includegraphics[width=\textwidth]{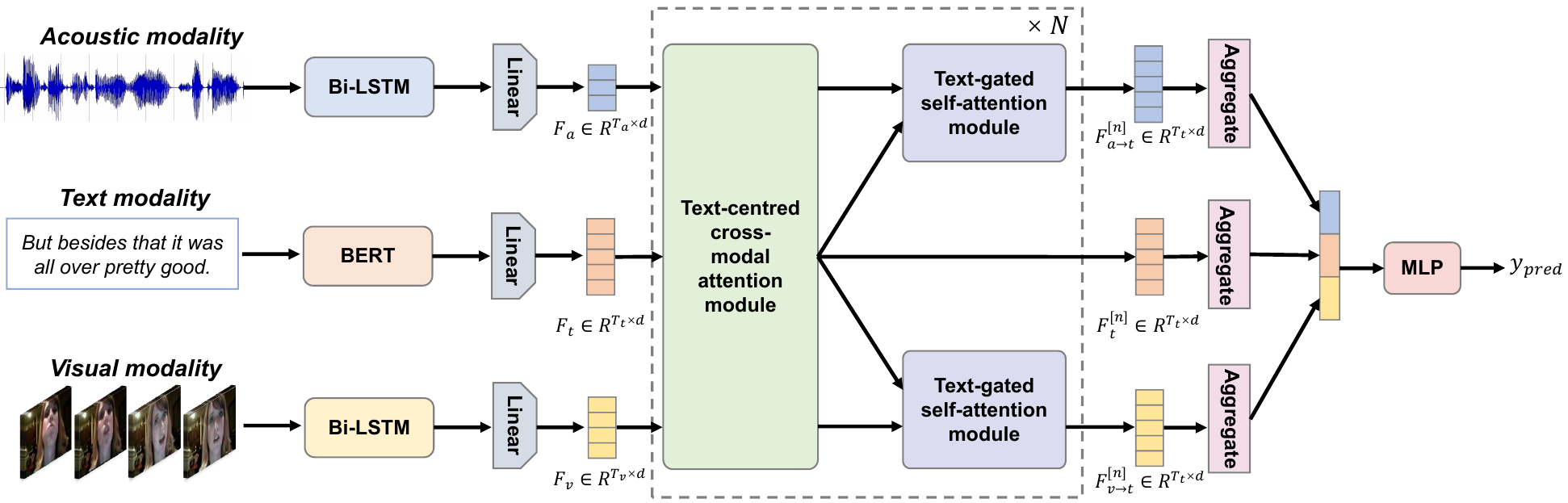}
\caption{The overall architecture of the proposed model TMRN.} \label{fig1}
\end{figure}\noindent In this work, the multimodal sentiment analysis task focuses on using the same video clip from the text (\textit{t}), visual (\textit{v}), and acoustic (\textit{a}) modalities as inputs to the model, which is represented as $ X_{m} \in R^{T_{m} \times d_{m}} $ for each modality  $m \in \left\{ {t, v, a} \right\}$.  For the rest of the paper,  $ T_{m} $ and  $d_{m}$  are used to represent sequence length and feature dimension of modality $m$, respectively. The goal of our model is to fully explore and fuse sentiment-related information from these input unaligned multimodal sequences to obtain a text-driven multimodal representation and thus predict the final sentiment analysis results.

\subsection{Overall Architecture}
\noindent The overall architecture of our TMRN is shown in Fig.~\ref{fig1}, which consists of three main components: 1) \textit{Unimodal feature extraction module}: we utilize pre-trained BERT \cite{devlin2018bert} to generate the extravagant representation of input words and process visual and acoustic features with Bi-LSTM \cite{hochreiter1997long}; 2) \textit{Modality reinforcement}: this part is composed of cross-stacking TCCA and TGSA modules to interact and reinforce the features. We divide the features into visual-text and acoustic-text pairs for cross-attention with the text modality as the query, while self-attention is performed on the text modality. Then, we fuse the pairs with an adaptive fusion mechanism. After that, we use the text modality as a gate to adding prior knowledge to the process of self-reinforcement of visual/acoustic modalities; 3) \textit{Fusion and output module}: we aggregate the final two-dimension features into one-dimension vectors and concatenate them for the downstream tasks. Our aim is to further guide and interact with acoustic and visual modalities through the text modality to obtain a text-dominated implicitly aligned fusion feature.
\begin{figure}
\centering 
\includegraphics[width=0.6\columnwidth]{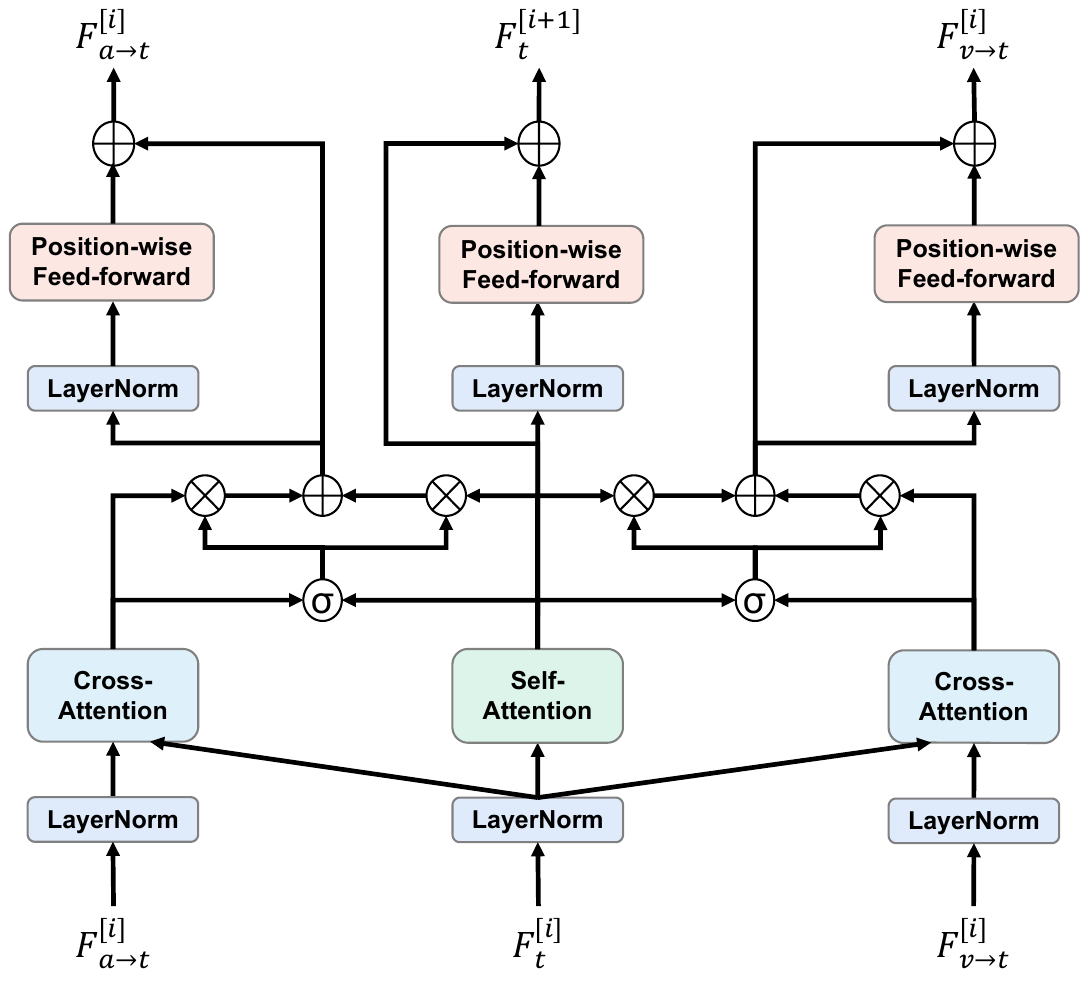}
\caption{The architecture of the Text-Centred Cross-modal Attention (TCCA) module.} \label{fig2}
\vspace{-8pt}
\end{figure}

\subsubsection{Unimodal Feature Extraction.} To obtain a stronger feature representation of the text, we use a pre-trained BERT \cite{devlin2018bert} model to extract the feature of the sentences:
\begin{equation}
F_{t~} = BERT\left( X_{t};\theta_{t}^{BERT} \right) \in R^{T_{t} \times d_{t}}.
\end{equation}
In acoustic and visual modalities, following \cite{zadeh2017tensor,yu2020ch}, we use pre-trained ToolKits to extract the initial features $X_{m}$ from raw data. Then, we use Bi-directional Long Short-Term Memory (BiLSTM)\cite{hochreiter1997long} to capture the timing characteristics:
\begin{equation}
F_{a} = BiLSTM\left( X_{a};\theta_{a}^{LSTM} \right) \in R^{T_{a} \times d_{a}},
\end{equation}
\begin{equation}
    F_{v~} = BiLSTM\left( X_{v};\theta_{v}^{LSTM} \right) \in R^{T_{v} \times d_{v}}.
\end{equation}

For subsequent calculations, we use one fully connected layer to project the features into a fixed dimension as $F_{m} \in R^{T_{m} \times d}$, where $m \in \left\{ t, a, v \right\}$.
\subsubsection{Modality Reinforcement.} This part includes two key modules: a Text-Centred Cross-modal Attention (TCCA) module and a Text-Gated Self-Attention (TGSA) module.
The architecture of TCCA is shown in Fig.~\ref{fig2}. Unlike \cite{lv2021progressive}, the visual and acoustic modalities share the same text self-attention block to reduce the amount of computation in our TCCA module. This unit is composed of two cross-attention blocks and one self-attention block. The cross-attention block takes $F_{t}^{\lbrack i\rbrack}$ and  $F_{m\rightarrow t}^{\lbrack i\rbrack}$ as its inputs, where $m \in \left\{ {a,v} \right\}$, and the superscript $\lbrack i\rbrack$ indicates the $i$-th modality reinforcement processes. First, we perform a layer normalization ($LN$) on the features like $
F_{m\rightarrow t}^{\lbrack i\rbrack} = LN\left( F_{m\rightarrow t}^{\lbrack i\rbrack} \right)$ and $F_{t}^{\lbrack i\rbrack} = LN\left( F_{t}^{\lbrack i\rbrack} \right)$, and then we put them into a Cross-Attention ($CA$) block:

\begin{equation}
       \begin{aligned}
           F_{m\rightarrow t}^{\lbrack i\rbrack} &= ~{CA}_{m\rightarrow t}^{\lbrack                i\rbrack}\left( {F_{m\rightarrow t}^{\lbrack i\rbrack},F_{t}^{\lbrack                i\rbrack}} \right),\\
            &= softmax\left( \frac{F_{t}^{\lbrack i\rbrack}W_{Q_{t}}W_{K_{m}}^{T}                {F_{m\rightarrow t}^{\lbrack i\rbrack}}^{T}}{\sqrt{d}} \right)F_{m\rightarrow                       t}^{\lbrack i\rbrack}W_{V_{m}},
       \end{aligned}
\end{equation}
where  $F_{m\rightarrow t}^{\lbrack 0\rbrack} = F_{m} \in R^{T_{m} \times d}$ and $F_{t}^{\lbrack 0\rbrack} = F_{t} \in R^{T_{t} \times d}$. Note that the sequence length of  $F_{m\rightarrow t}^{\lbrack i\rbrack}$ is updated to $T_t$  after the first $CA$ block. And the Self-Attention ($SA$) block takes $
F_{t}^{\lbrack i\rbrack}$ as input to obtain $F_{t}^{\lbrack{i + 1}\rbrack} \in R^{T_{t} \times d}$:
\begin{equation}
    \begin{aligned}
        F_{t}^{\lbrack{i + 1}\rbrack} &= ~{SA}_{t}^{\lbrack i\rbrack}\left( F_{t}^{\lbrack i\rbrack} \right),\\ &= softmax\left( \frac{F_{t}^{\lbrack i\rbrack}W_{Q_{t}}W_{K_{t}}^{T}{F_{t}^{\lbrack i\rbrack}}^{T}}{\sqrt{d}} \right)F_{t}^{\lbrack i\rbrack}W_{V_{t}}.
    \end{aligned}
\end{equation}
\begin{figure}[t]
\centering 
\includegraphics[width=0.7\columnwidth]{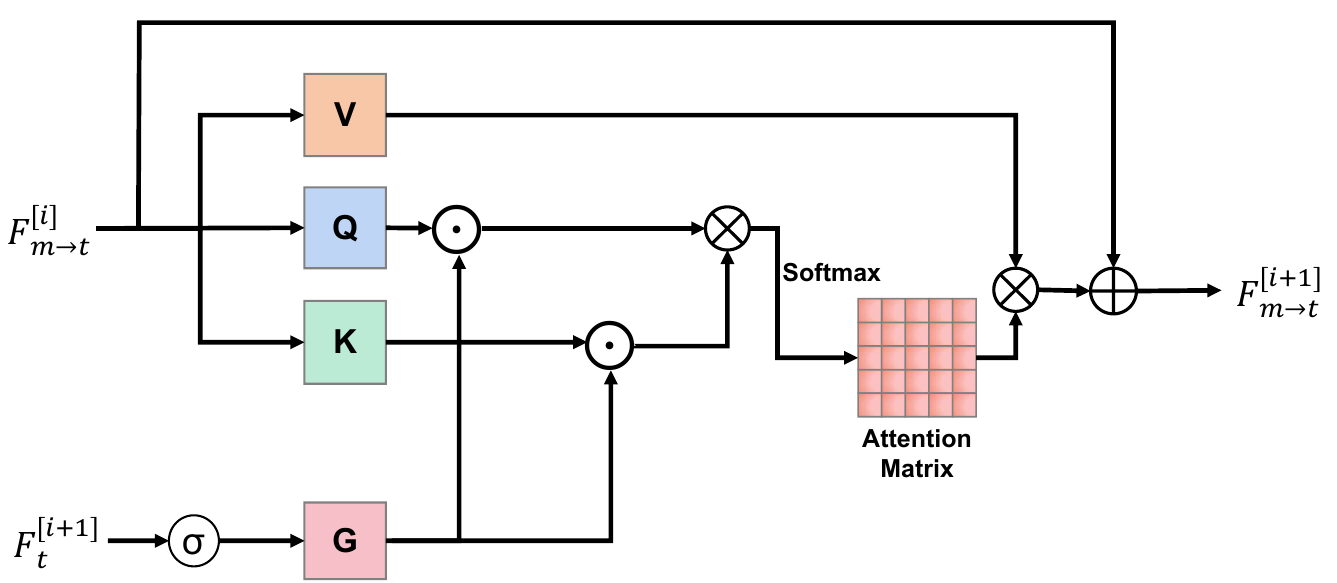}
\caption{The architecture of the Text-Gated Self-Attention (TGSA) module.} 
\label{fig3}
\end{figure}
Then the reinforced features $F_{t}^{\lbrack{i + 1}\rbrack}$ and $F_{m\rightarrow t}^{\lbrack i\rbrack}$ are processed via the following adaptive fusion mechanism:
\begin{equation}
G^{\lbrack i\rbrack} = \sigma\left( F_{t}^{\lbrack{i + 1}\rbrack}*W_{t}^{\lbrack i\rbrack} + F_{m\rightarrow t}^{\lbrack i\rbrack}*W_{m\rightarrow t}^{\lbrack i\rbrack} + b^{\lbrack i\rbrack} \right),
\end{equation}
\begin{equation}
F_{m\rightarrow t}^{\lbrack i\rbrack} = G^{\lbrack i\rbrack} \odot F_{t}^{\lbrack{i + 1}\rbrack} + \left( {1 - G^{\lbrack i\rbrack}} \right) \odot F_{m\rightarrow t}^{\lbrack i\rbrack},
\end{equation} 
where $\sigma$ denotes the sigmoid non-linearity function, $\odot$ denotes element-wise multiplication. We can determine the passed proportions of $F_{t}^{\lbrack{i + 1}\rbrack}$ and $F_{m\rightarrow t}^{\lbrack{i}\rbrack}$ via the learnable parameters $W_{t}^{\lbrack i\rbrack}$, $W_{m\rightarrow t}^{\lbrack i\rbrack}$, and $
b^{\lbrack i\rbrack}$. This operation can filter the incorrect information produced by the cross-modal interactions, and measure the fusion ratio of two modalities. After that, we process $F_{t}^{\lbrack{i + 1}\rbrack}$ and $F_{m\rightarrow t}^{\lbrack i\rbrack}$ by a Position-wise Feed-Forward layer ($PFF$) with skip connection, as in the Transformer\cite{vaswani2017attention}:
\begin{equation}
    F_{m\rightarrow t}^{\lbrack i\rbrack} = PFF\left( {LN\left( F_{m\rightarrow t}^{\lbrack i\rbrack} \right)} \right) + F_{m\rightarrow t}^{\lbrack i\rbrack},
\end{equation}
\begin{equation}
    F_{t}^{\lbrack{i + 1}\rbrack} = PFF\left( {LN\left( F_{t}^{\lbrack{i + 1}\rbrack} \right)} \right) + F_{t}^{\lbrack{i + 1}\rbrack}.
\end{equation}

After the TCCA module, we obtain unified dimensional features of three modalities.  We think that the relationships within each modality are complementary to the cross-modal relations, so we do self-attention for $F_{v\rightarrow t}^{\lbrack i\rbrack}$ and $
F_{a\rightarrow t}^{\lbrack i\rbrack}$, while using the $F_{t}^{\lbrack{i + 1}\rbrack}$ as a gate to activate or deactivate the corresponding key and value channels:
\begin{equation}
g^{\lbrack i\rbrack} = \sigma\left( Linear\left( F_{t}^{\lbrack{i + 1}\rbrack};\theta_{g} \right) \right),
\end{equation}
\begin{equation}
    {gF}_{m\rightarrow t}^{\lbrack i\rbrack} = \left( 1 + g^{\lbrack i\rbrack} \right){\odot F}_{m\rightarrow t}^{\lbrack i\rbrack}.
\end{equation}
The query and key from visual/acoustic modalities are then modulated by the gate from the text modality:
\begin{equation}
    \begin{aligned}
        F_{m\rightarrow t}^{\lbrack{i + 1}\rbrack} &= {TGSA}_{m}^{\lbrack i\rbrack}\left( {F_{m\rightarrow t}^{\lbrack i\rbrack},{~gF}_{m\rightarrow t}^{\lbrack i\rbrack}} \right) ,\\ &= softmax\left( \frac{{gF}_{m\rightarrow t}^{\lbrack i\rbrack}W_{Q_{m\rightarrow t}}W_{K_{m\rightarrow t}}^{T}{{gF}_{m\rightarrow t}^{\lbrack i\rbrack}}^{T}}{\sqrt{d}} \right)F_{m\rightarrow t}^{\lbrack i\rbrack}W_{V_{m\rightarrow t}} + F_{m\rightarrow t}^{\lbrack{i}\rbrack}.
    \end{aligned}
\end{equation}
The architecture of the TGSA is shown in Fig.~\ref{fig3}.
\subsubsection{Fusion and Output Module.} Here, we utilize a simple attention approach to aggregate the reinforced features of the three modalities. Specifically, given the feature $
F_{m}^{\lbrack n\rbrack} \in R^{T_{m} \times d}$ for modality $m$ output by the last TGSA module, we get the attention weight matrix:
\begin{equation}
    a_{m} = softmax\left( \frac{F_{m}^{\lbrack n\rbrack}W_{m}}{\sqrt{d}}\right)^{T}\in R^{1 \times T_{m}},
\end{equation}
where $W_{m} \in R^{ d}$ denotes the linear projection parameter, and $a_{m}$ denotes the attention weight matrix for the feature $F_{m}^{\lbrack n\rbrack}$. Then we aggregate the features with the attention weights:
\begin{equation}
f_{m} = a_{m}F_{m}^{\lbrack n\rbrack} \in R^{1\times d}.
\end{equation}

Eventually, we concatenate all of the three modalities’ features as $
f = \left\lbrack f_{t}; f_{a}; f_{v} \right\rbrack \in R^{1\times3d}$ as the fused feature passing through a Multi-Layer Perceptron ($MLP$) to make the final prediction $
y_{pred}$:
\begin{equation}
    y_{pred} = \Phi\left( f;\theta_{\Phi} \right),
\end{equation}
where the $\Phi (\cdot)$ is a $MLP$ parameterized  by $\theta_{\Phi}$.
\vspace{-10pt}
\section{Experiments}
\noindent In this section, we empirically evaluate our model on two datasets that are frequently used to benchmark the MSA task in prior works, and we introduce the datasets, implementation details, and the results of our experiments.
\subsection{Datasets and Evaluation Metrics}

\noindent MOSI\cite{zadeh2016mosi} dataset is a widely used benchmark dataset for the MSA task. It comprises 2,199 short monologue video clips taken from 93 Youtube movie review videos.  Its predetermined data partition has 1,284 samples in the training set, 229 in the validation set, and 686 in the testing set. MOSEI \cite{zadeh2018multimodal} dataset is an improvement over MOSI. It contains 22,856 annotated video segments (utterances) from 5,000 videos, 1,000 distinct speakers, and 250 different topics. Its predetermined data partition has 16,326 samples in the training set, 1,871 in the validation set, and 4,659 in the testing set. Each sample in both MOSI and MOSEI is manually annotated with a sentiment score between $[-3,3]$, which indicates the polarity and relative strength of expressed sentiment. 
The polarity is indicated by positive/negative, and strength is indicated by absolute value.
As in the previous works\cite{liang2021attention,lv2021progressive}, we evaluate the model performances by the 7-class accuracy ($Acc_7$), the binary accuracy ($Acc_2$), mean absolute error ($MAE$), the correlation of the model’s prediction with human ($Corr$), and the $F1$ score.
\begin{table}[t]
\centering
\caption{Comparison results on the MOSI. For $Acc_2$ and $F1$, we have two sets of non-negative/negative (left) and positive/negative (right) evaluation results. }\label{tab1}
\begin{tabular}{lccccc}
\toprule
Method	& $MAE \downarrow$ & $Corr \uparrow$ & $Acc_7 \uparrow$ & $Acc_2  \uparrow$ & $F1  \uparrow$
\\
\midrule
TFN &  0.901 & 0.698 &34.9 &-/80.8 & -/80.7\\[1 ex]
LMF &  0.917 & 0.695& 33.2& -/82.5&-/82.4 \\[1 ex]
MulT &  0.861 & 0.711& - &81.5/84.1 &80.6/83.9 \\[1 ex]
MISA &  0.783 & 0.761 & 42.3&81.8/83.4 & 81.7/83.6\\[1 ex]
MAG-BERT & 0.731 & 0.789 & - & 82.5/84.3 & 82.6/84.3 \\[1 ex]
Self-MM & 0.718 & \textbf{0.796} &  46.04&  82.62/84.45& 82.55/84.44\\[1 ex]
\textbf{TMRN(ours)} & \textbf{0.704} & 0.784 & \textbf{48.68} & \textbf{83.67/85.67} & \textbf{83.45/85.52}\\
\bottomrule
\end{tabular}
\end{table}

\begin{table}[t]
\centering
\caption{Comparison results on the MOSEI.}\label{tab2}
\begin{tabular}{lccccc}
\toprule
Method	& $MAE \downarrow$ & $Corr \uparrow$ & $Acc_7 \uparrow$ & $Acc_2  \uparrow$ & $F1  \uparrow$
\\
\midrule
TFN &  0.593 & 0.700 & 50.2 & -/82.5 & -/82.1\\[1 ex]
LMF &  0.623 & 0.677& 48.0& -/82.0&-/82.1 \\[1 ex]
MulT &  0.580 & 0.703& - & 82.5 & -/82.9 \\[1 ex]
MISA &  0.568 & 0.724 & - &82.59/84.23 & 82.67/83.97\\[1 ex]
MAG-BERT & 0.539 & 0.753 & - & 83.8/85.2 & 83.7/85.1 \\[1 ex]
Self-MM & 0.536 & \textbf{0.763} & \textbf{54.5} & 82.59/84.95 & 82.9/84.85\\[1 ex]
\textbf{TMRN(ours}) & \textbf{0.535} & 0.762 & 53.65 & \textbf{83.39/86.19} & \textbf{83.67/86.08}\\
\bottomrule
\end{tabular}
\vspace{-6pt}
\end{table}

\begin{table}
\centering
\caption{Ablation results of our TMRN on the MOSI.}\label{tab3}
\begin{tabular}{lccccc}
\toprule
Model	& $MAE \downarrow$ & $Corr \uparrow$ & $Acc_7 \uparrow$ & $Acc_2  \uparrow$ & $F1  \uparrow$
\\
\midrule
\textbf{Full method} &   \textbf{0.7041} & \textbf{0.7844} & \textbf{48.68} & \textbf{83.67/85.67} & \textbf{83.45/85.52} \\[1 ex]
w/o A &  0.8114 & 0.7426 & 45.48&81.05/81.86 & 81.09/81.96\\
w/o V &  0.8452 & 0.7382 &41.69 &80.61/81.71 & 80.67/81.82\\\midrule
Acoustic-oriented &  0.7508 & 0.7658 & 43.00&81.92/83.38 &81.85/83.36 \\
Visual-oriented &  0.7956 & 0.7309 & 41.69&82.07/83.23 &82.06/83.27 \\ \midrule
w/o TCCA & 0.7498 & 0.7817 &44.75 &83.09/84.76 &83.03/84.75\\
w/o TGSA & 0.7467 & 0.7824 &45.33 &80.45/81.71 &80.50/81.79\\
\bottomrule
\end{tabular}
\vspace{-6pt}
\end{table}
\subsection{Implementation Details}
\noindent All models are built on the Pytorch toolbox \cite{paszke2019pytorch} with two Quadro RTX 8000 GPUs. The Adam optimizer\cite{kingma2014adam} is adopted for network optimization. For the MOSI and MOSEI datasets, the training setting follows: the batch sizes are $\{128, 64\}$, the epochs are $\{100, 40\}$, the learning rates are $
\{{1e}^{- 3}, {2e}^{- 3}\}$, and the hidden dimension \textit{d} is 128. The number of TCCA and TGSA is $N =3$. 

\subsection{Comparison with State-of-the-Art Methods}
\noindent The proposed approach is compared to the existing state-of-the-art (SOTA) baselines, including TFN \cite{zadeh2017tensor}, LMF \cite{liu2018efficient}, Mult \cite{tsai2019multimodal}, MISA \cite{hazarika2020misa}, MAG-BERT \cite{rahman2020integrating}, and Self-MM \cite{yu2021learning}. Table~\ref{tab1} and~\ref{tab2} show the comparison results on the MOSI and MOSEI, respectively.  
The result of Self-MM \cite{yu2021learning} is reproduced from open-source code with hyper-parameters provided in the original paper. 

The proposed TMRN significantly outperforms most previous methods~\cite{zadeh2017tensor,liu2018efficient,tsai2019multimodal,hazarika2020misa,rahman2020integrating} by considerable margins on all metrics in both datasets, demonstrating the superiority of our method.
In addition, our model is superior to the current SOTA Self-MM~\cite{yu2021learning} in most metrics ($i.e.,$ $MAE, Acc_7, Acc_2, F1$ scores on the MOSI, and $MAE, Acc_2, F1$ scores on the MOSEI.) with better or competitive performance, suggesting the effectiveness of our text-oriented design philosophy.

\subsection{Ablation Study}
\noindent The overall performance has proven the superiority of TMRN. To understand the necessity of the different components and the dominance of the text modality, we conduct systematic ablation experiments on the MOSI, as shown in Table~\ref{tab3}.

\subsubsection{Importance of Modality.} We remove a modality separately to explore the performance of our model. 
Both declining results indicate the importance of visual and acoustic modalities when removing the visual or acoustic sequences.
Furthermore, the performance degradation is more severe when the visual modality is removed. This is in line with the previous work~\cite{yang2022disentangled}. This result suggests that the information in nonverbal modalities complements the text modality.

\subsubsection{Importance of  Center Modality.} To demonstrate the dominance of the text modality, we replace the other two modalities as the dominant modality for the experiments. 
The acoustic- and visual-oriented models invariably suffer from significant performance degradation.
These observations demonstrate that the text modality is richer in semantics and less noisy, which leads to better feature reinforcement of the other two modalities.

\subsubsection{Importance of Module.} Finally, we explore the importance of the proposed components by removing the TCCA and TGSA modules separately. For the TCCA module, we remove the cross-attention block and only do self-attention for text modality. We can see that the gain degrades when removing one of the modules. These observations suggest that adequate guidance of the text modality is necessary and indispensable.
\begin{figure}[t]
\centering 
\includegraphics[width=0.85\linewidth ]{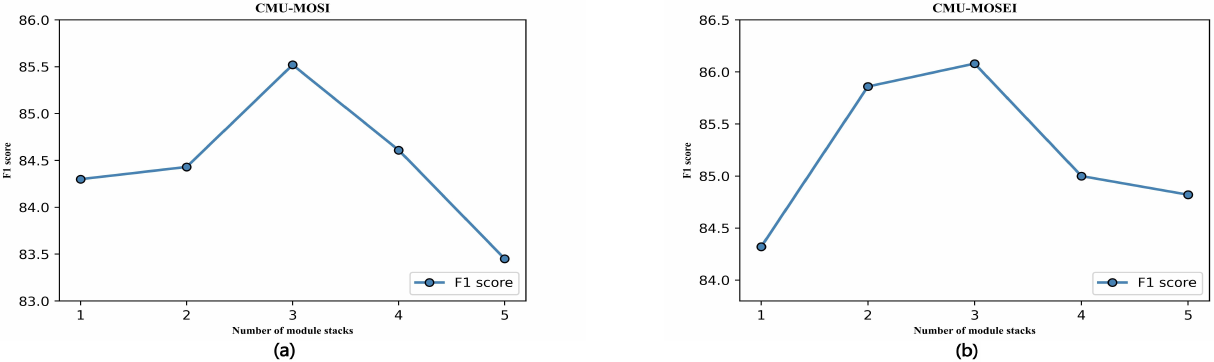}
\caption{Performance of TMRN with different parameter $N$ on MOSI and MOSEI. } 
\label{fig4}
\end{figure}
\subsection{Sensitivity of Parameter} 
\noindent In order to explore the effect of parameter $N$ on the model performance, we conducted experiments on MOSI and MOSEI datasets with different parameters $N$. The results are summarized in Fig. \ref{fig4}.
With the increase of $N$, we find that the $F1$ scores show a trend of increasing and then decreasing, and the network performs best when $N=3$. In our conjecture, the larger $N$  can result in better modality reinforcement. However, experiments show us that too many layers may bottleneck the ability of the text modality to guide the other two modalities. We should choose the appropriate network for different datasets, which is exactly what our proposed TMRN can flexibly do. If migrating our model to a more complex dataset, we can properly increase the number of TCCA and TGSA modules to achieve the best performance.
\section{Conclusion}
\noindent This paper presents a text-oriented multimodal sequence reinforcement network to achieve interaction and fusion over unaligned sequences of three modalities in the context of multimodal human sentiment analysis. The work is based on inter- and intra-modal attention mechanisms, and the attention of the other two modalities is guided throughout by sequences from the text modality, enabling the alternate transfer of information within and across modalities. We experimentally observe that our approach can achieve better performance than the baselines in MSA benchmarks.

\subsubsection{Acknowledgments} This work was supported in part by the National Key R\&D Program of China (2021ZD0113503), and in part by the Shanghai Municipal Science and Technology Major Project (2021SHZDZX0103).
\bibliographystyle{splncs04}
\bibliography{paper.bib}

\end{CJK}

\end{document}